# ELECTROMAGNETIC E2 TRANSITION PROBABILITIES IN $^{120}$Xe AND $^{118}$Te - $N$=66 NUCLEI

A.A. Pasternak[a], A.D. Efimov[a], E.O. Podsvirova[a], V.M. Mikhajlov[b], J. Srebrny[c], T. Morek[c], Ch. Droste[c], Y. Sasaki[d], M. Oshima[e], S. Juutinen[f], and G.B. Hagemann[g]

[a] A.F. Ioffe Physical-Technical Institute, 194021, St.-Petersburg, Russia
[b] Physical Institute of St.-Petersburg State University, Russia
[c] Nuclear Physics Division, IEP, Warsaw University, Poland
[d] Tandem Accelerator Center, University of Tsukuba, Japan
[e] Japan Atomic Energy Research Institute, Tokai-mura, Japan
[f] Department of Physics, University of Jyväskylä, Finland
[g] The Niels Bohr Institute, Copenhagen, Denmark

Lifetimes of the yrast states in $^{120}$Xe and ground state band below and above band crossing in $^{118}$Te have been measured by DSAM in the $^{111}$Cd($^{12}$C,3n) reaction and by DSAM and RDM in the $^{109}$Ag($^{13}$C,p3n) reaction, respectively. The results are compared with E2 transition probabilities in $^{119}$I and $^{128}$Ba. The experimental data are compared with calculation done in the framework of the IBM1 model in the O(6) and SU(5) limits.

## 1. Introduction

The motivation for the lifetime study of $^{120}$Xe [1] and $^{118}$Te [2] was our previous investigation of $^{119}$I where lifetime of 60 levels in 10 bands have been measured [3, 4]. The energy structure of the low-lying part of the yrast decoupled negative parity band built on the $h_{11/2}$ state in this nucleus is very similar to the yrast state band below the backbending in the neighbour $^{120}$Xe nucleus (Fig. 1). This figure also shows that the E2 $\Delta I$=2 transitions in the $g_{9/2}$ band have energies similar to the corresponding E2 transitions in the neighbouring $^{118}$Te nuclei. The $B$(E2) values for the $h_{11/2}$ and $g_{9/2}$ bands of $^{119}$I differ from each other [4], therefore, a comparison with corresponding values in $^{118}$Te and $^{120}$Xe can be helpfull to explain this fact. Before this work, only a few experimental lifetime values in $^{120}$Xe and $^{118}$Te nuclei were known [5, 6, 7, 8]. For low-lying ($I < 8$) exited states of $^{118}$Te there were not any lifetime data.

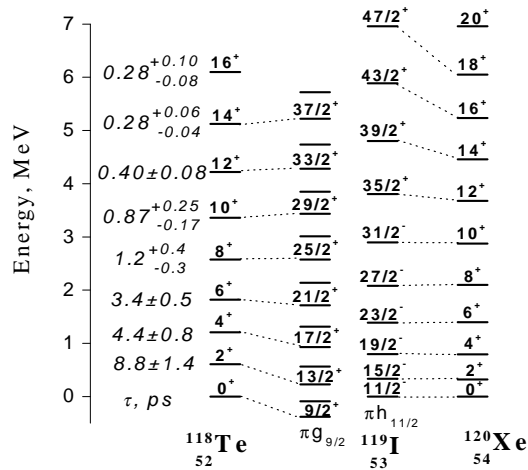

Fig. 1. Ground state band levels in $^{118}$Te and $^{120}$Xe in comparison with $\pi g_{9/2}$ and $\pi h_{11/2}$ bands of $^{119}$I

## 2. Experiment

Lifetimes in $^{120}$Xe have been measured by the Doppler Shift Attenuation method (DSA) in the $^{111}$Cd($^{12}$C,3n) reaction at a beam energy of $E = 56$ MeV [1]. The experiment was performed at the JAERI Tandem Accelerator (Japan). The $\gamma\gamma$ coincidences were collected by the GEMINI array. A thick target (30 mg/cm$^2$ metallic foil) has been used.

Lifetimes in $^{118}$Te have been determined with the DSA and Recoil Distance (RD) methods using the $^{109}$Ag($^{13}$C,p3n) reaction at $E = 54$ MeV [2]. The experiment was performed at TAL NBI (Denmark). The $\gamma\gamma$ coincidences were collected by the NORDBALL array. For the DSA method a target of 5.7 mg/cm$^2$ has been used. RD measurements have been done using a self-supporting 0.82 mg/cm$^2$ target [3].

The analysis of experimental $\gamma$-lineshapes was carried out using an updated version of Monte-Carlo codes COMPA, GAMMA, SHAPE [3, 10]. The main features of the used approach are the following:

a. The kinematical spread of the initial recoils is calculated on the basis of a statistical model taking into account step by step the evaporation of light particles from compound nucleus.

b. The slowing down and multiple scattering of the recoils can be calculated for several stopping layers. Recoil Distance spectra are regarded as a special case of DSA with three layers- target,vacuum and stoper. More details concerning the Recoil Distance Doppler Shift Attenuation (RDDSA) method can be found in Refs [3,11,12]).

c. The number, solid angles and arrangement of $\gamma$-detectors are not limited.

d. The number of levels and branches of cascade feeding is not limited. Sidefeeding cascades are included into the Monte-Carlo simulation from each state to the level of interest. Any condition of $\gamma$-$\gamma$-coincidence gating ("above", "below" etc.) can be taken into account by the Monte-Carlo techniques.

e. Overlapping Doppler broadened lines can be analysed using lifetimes as lineshape parameters.

The Lindhard correction factors for electronic ($f_e$) and nuclear ($f_n$) components of the stopping power of the recoils have been measured by lineshape analysis using the "semi-thick target" method [10, 14] and values of $f_e = 1.27 \pm 0.07$ and $f_n = 0.77 \pm 0.07$ have been obtained for the case of the $^{119}$I recoils moving in the $^{109}$Ag target [3, 13]. It follows from our measurements that, for recoil velocities $v = 1\%c$ the electronic stopping power taken from the tables of Ziegler et al. [19] ( computer codes TRIM-95, SRIM-2000.) can be two times smaller than the experimental values, whereas the nuclear stopping power is close the predicted one.

To extract the lifetime of high spin states the sum of spectra gated below the line of interest was used. The sidefeeding effective time $\tau_{sf}$ is expected to be small in both nuclei [1, 2]. From precise lineshape analysis of the 819 keV $18^+ \rightarrow 16^+$ transition in $^{120}$Xe the value of $\tau_{sf} = 0.040 \pm 0.015$ ps has been evaluated .

For the analysis of the high spin levels in $^{118}$Te sum of the spectra taken from all 20 NORDBALL detectors have been used. Symmetrical Doppler broadening $\gamma$ lines corresponding to relatively short lived levels obtained in this way were analyzed [2]. In the case of $^{120}$Xe the DSA lineshape analysis of spectra gated on the flight component of the 819 keV $18^+ \rightarrow 16^+$ $\gamma$ line as well as on $\gamma$ lines below the $10^+$ state has been done. As a result, the lifetimes of the $16^+$ and $10^+$ levels from the unresolved 773 keV doublet consisting of the $16^+ \rightarrow 14^+$ and $10^+ \rightarrow 8^+$ transitions have been extracted [1].

Lifetimes of the 2+, 4+ and 6+ levels in $^{118}$Te have been evaluated from the RD spectra gated on the flight component of the 753 keV $8^+ \rightarrow 6^+$ line. Since the 600, 606 and 615 keV lines, corresponding to the $4^+ \rightarrow 2^+$, $2^+ \rightarrow 0^+$ and $6^+ \rightarrow 4^+$ transitions are overlapping each other, $\tau$ values were determined by the RDDSA method [2]. (Fig. 2).

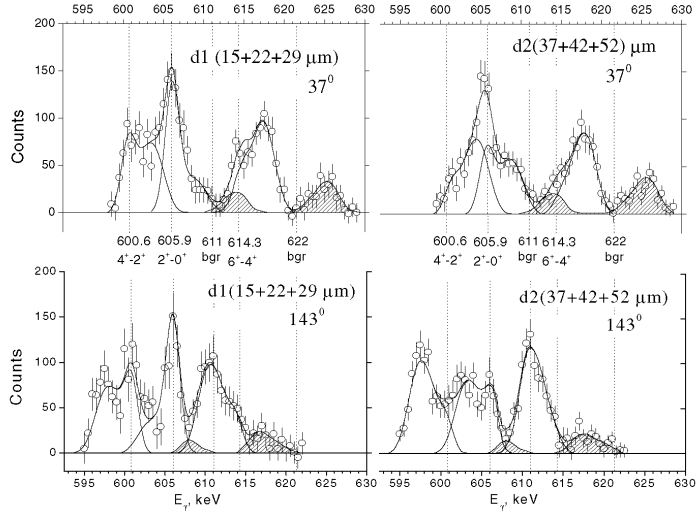

Fig. 2. Analysis of the multiplet consisted of the 600, 605, 611, 614 and 622 keV lines. "bcg" - background lines, d1, d2 - spectra being the sum of spectra obtained in the RD method for the target-stoper distances given in brackets.

## 3. Discussion

One of the first attempts of describing the excitation energies and $B(E2)$ values in the backbending region has been done using model based on the IBM1 model, which involved high-spin phonons in addition to s- and d-bosons [15]. This approach has been improved by using microscopic calculations and applied to $^{110}$Cd and $^{126}$Ba [16, 9]. A detailed description of this model, named IBM+2qp, is presented in [9]. The last developments of this model applied to $^{118}$Te and $^{120}$Xe are presented in Refs [1, 2, 18]. In the left panel of Fig. 3 a comparison of the experimental $B(E2)$ values in $^{120}$Xe with data from the recently investigated $^{128}$Ba nucleus [17] is presented. It is worth adding that lifetime obtained for $^{118}$Te give us the unique possibility to analyse $B(E2)$ values in the collective ground state band above the band crossing. It is easy to see that in both nuclei corresponding O(6) limits overestimate the experimental $B(E2)$ values near the backbending region.

It can be seen in Fig. 1 and Fig. 3 (left panel) that the decoupled $\pi h_{11/2}$ band in $^{119}$I is similar to the yrast band of $^{120}$Xe both in the excitation energies and in the B(E2) dependence on the spin. The collectivity of the $\pi$h$_{11/2}$ band is even higher than that of the yrast band in $^{120}$Xe. It can be due to core polarization effects. The situation for the strongly coupled g$_{9/2}$ band in $^{119}$I turns out to be more complex. Fig. 3 (right panel) shows that E2 transition probabilities for the g$_{9/2}$ band in $^{119}$I are significantly smaller than the corresponding values for the low spin levels of $^{120}$Xe and even $^{118}$Te. Moreover there is a difference in the spin-dependence of the $B$(E2) values for $^{119}$I and $^{118}$Te. Fig. 3 (right panel) shows that the $B$(E2) values for the ground state band in $^{118}$Te increases with spin in contradiction with theoretical results of IBM1 model in O(6) limit but are close to SU(5) limit. It is of interest that in $^{119}$I above backbending the $B$(E2) values and their spin-dependence drastically differ from the low-spin region of the $^{119}$I band but are similar to those of the low spin states of $^{118}$Te. The low collectivity of the g$_{9/2}$ band below backbending has to be accounted for and this is a challenge to the theory.

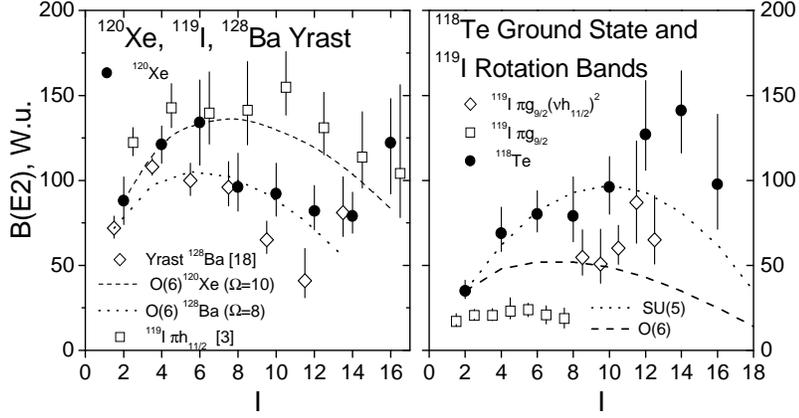

Fig. 3. Experimental spin dependence of $B$(E2) in $^{120}$Xe, $^{128}$Ba, $^{119}$I (left panel) and $^{118}$Te, $^{119}$I (right panel). The spin values for bands containing the $\pi$h$_{11/2}$ and $\pi$g$_{9/2}$ orbitals are shifted by spin value of their bandheads

## 4. Summary

The collectivity of the $\pi$h$_{11/2}$ decoupled band in $^{119}$I is similar to the $^{120}$Xe core while B(E2) values in the $\pi$g$_{9/2}$ rotational band of $^{119}$I are drastically smaller even in comparison with the relatively weak collectivity of the $^{118}$Te ground state band. The similar dependence of $B$(E2) values vs. spin has been found for the $^{120}$Xe and $^{128}$Ba nuclei.

We would like to thank S.G. Rohoziński and R.M. Lieder for enlightening discussions and critical remarks.